\documentclass{aastex}
\usepackage{spr-astr-addons}
\usepackage{url}
\usepackage{adjustbox}
\usepackage{array}
\usepackage{verbatim}
\usepackage{stmaryrd}
\usepackage{epsfig}
\usepackage{epstopdf}
\usepackage{multirow}
\usepackage{booktabs}
\usepackage{rotating}
\usepackage{etoolbox}
\usepackage{boondox-cal}

\newcolumntype{L}[1]{>{\raggedright\let\newline\\\arraybackslash\hspace{0pt}}m{#1}}
\newcolumntype{C}[1]{>{\centering\let\newline\\\arraybackslash\hspace{0pt}}m{#1}}
\newcolumntype{R}[1]{>{\raggedleft\let\newline\\\arraybackslash\hspace{0pt}}m{#1}}

\RequirePackage{color}

\begin{document}
	
	\title{$\beta$-decay properties of some astrophysically important Sc-isotopes}
	\shorttitle{$\beta$-decay properties of Sc isotopes}
	\shortauthors{Author et al.}
	
	\author{Fakeha Farooq\altaffilmark{1}} \and \author{Jameel-Un Nabi\altaffilmark{2,3}} \and
	\author{Ramoona Shehzadi\altaffilmark{1}}
	
	\email{ramoona.physics@pu.edu.pk}
	
	\altaffiltext{1}{Department of Physics, University of the Punjab,
		Lahore, Pakistan.} 
	\altaffiltext{2}{University of Wah, Quaid Avenue, Wah Cantt 47040, Punjab, Pakistan.}
	\altaffiltext{3}{Faculty of Engineering Sciences\\GIK Institute of Engineering Sciences and Technology, Topi
		23640, Khyber Pakhtunkhwa, Pakistan.}
	\altaffiltext{1}{Corresponding author email: ramoona.physics@pu.edu.pk}

\begin{abstract}
In late progressive-stages of heavy stars, electron capture and $\beta^{\pm}$-decay
are the governing processes. The weak rates are essential inputs for the modeling
of the stages of high-mass stars before supernova explosions. As per results
obtained from previous simulations, weak rates of Scandium isotopes contribute
substantially in changing the lepton-to-baryon ratio (Y$_{e}$) of the
nuclear matter of core. In the present analysis, we have reported some important
$\beta$-decay properties of crucial Sc isotopes in astrophysical environment
having 49 $\le$ A $\le$ 54. The investigation includes GT-strength distributions,
terrestrial half-lives and stellar rates of electron capture (EC) and $\beta^{-}$-decay reactions.
The calculations are performed in a microscopic way by using the proton-neutron (pn)
quasi-particle random phase approximation (QRPA) model over wide temperature
($10^{7} - 3 \times 10^{10}$)\;K and density ($10-10^{11}$)\;g/cm$^{3}$ domains.
In addition, a comparison of our calculated results have been done
with experimental and theoretical data, where available. A good agreement
between our half-lives calculations and experimentally measured results is observed.
Our calculated weak $\beta^{-}$-decay and EC rates have been compared
with earlier computed rates of Independent-Particle Model (IPM) and
Large-Scale Shell Model (LSSM). At high stellar temperature and density
conditions, our calculated rates of $\beta^{-}$-decay are smaller than those from the
other models. The decrement in our rates approaches up to 1-3 orders of magnitude.
In contrast, our EC rates are larger at high temperature by
up to 1-2 orders of magnitude.
\end{abstract}
\section{Declarations}
\label{sec:declare}
\subsection{Funding}
Not applicable.
\subsection{Conflicts of interest/Competing interests}
The authors do not have a conflict of interest with anyone to the best of their knowledge.
\subsection{Availability of data and material}
The calculated data and relevant material may be requested from corresponding author.
\subsection{Code availability}
Self-written code.
\section{Introduction}
\label{sec:intro} In astrophysical environments, the weak nuclear force has a pronounced
effect during the pre- and post-phases of core-collapse in high-mass stars
~\citep{Burbidge57, Bethe79}. These interactions unfold the underlying scenarios involved
during main sequence hydrogen-burning, hydrostatically equilibrated iron-core during pre-collapse
stage~\citep{Arn77, Lang15}, late evolutionary phases of stars, and thermonuclear and gravitational
collapse of core~\citep{Iwa99, Brac00, Heger01, Hix03, Jan07}. The processes mediated by
weak-force, namely; $\beta^{\pm}$-decays and electron captures (EC), alter the nucleosynthsis yield of
different exotic nuclei in the phases of core-collapse and explosive burnings. In addition,
$\beta^{-}$-decays and EC compete with each other and hence change Y$_{e}$ (lepton-to-baryon fraction) in
stellar-core composition~\citep{Burbidge57, Aufder94b, Wall97, Fuller}.

In type II supernova, when core mass surpasses the appropriate
limit of Chandrasekhar mass, it enters into a collapsing stage, where EC on iron-peak nuclei results in
 reduction of electron pressure and core energy.
This reduction in energy is linked with the production of neutrinos (anti-neutrinos) in EC ($\beta^{-}$-decay)
which leave the stars~\citep{Lang99}. Thereupon, with the decrease in Y$_{e}$ the nuclear matter
turns into neutron rich environment, where $\beta^{-}$-decay becomes a dominant weak process and starts
competing with EC~\citep{Lang00, Mart00}. Therefore, $\beta^{-}$-decay and EC processes and their
corresponding rates are important nuclear inputs for a realistic simulations of astrophysical environments.
The determination of these reactions and their rates substantially depend upon Gamow-Teller (GT-) resonance
~\citep{Bethe79}. GT-transitions and knowledge of energy distributions in these strengths give deep insight
to the nuclear structure~\citep{Ost92, Fuj11, Sar18, Sax18}.
Therefore, an accurate and detailed knowledge of GT-strength distributions B(GT$_{-}$)/B(GT$_{+}$) is essential for
$\beta^{-}$/EC rates.

In the first place, \citet{Fuller} (FFN) estimated stellar weak rates
with the systematic inclusion of GT-resonance based upon the Independent Particle Model (IPM). However, in charge exchange (CE)
experiments~\citep{Kateb94, Ron93, Rap83, And90, Good80, Alford90, Williams90, Alford93} a quenching
of total strength in GT-distribution has been observed in comparison to the IPM calculated strength. More significant results
of these CE experiments were the fragmentation of the GT-strengths over many daughter-nucleus
levels. Later,~\cite{Aufder94b} extended the FFN work with the inclusion of
GT quenching. However, in the calculations done by both of the groups, the GT-centroid was not correctly
placed. Afterwards, in 1996, ~\citet{Aufder96} highlighted the flaw in the parametrization
used in their previous work and in the FFN calculations as well.

The CE experiments provide robust benchmark for GT-strengths. Although,
the technologies and methods used in modern laboratories for nuclear measurements
have become tremendously advanced, it is still challenging to measure most of the energy levels
and matrix elements for nuclear transitions. One is constrained to perform calculations of weak decay rates under stellar conditions. A full blown microscopic nuclear model to preform these calculations is in order. Under
potentially high temperature of central core material of stars, the Fermi energies of electrons also imply the contributions of
excited states of parent nuclei to the total weak rates~\citep{Bahcall64, Bahcall74, FulMe91}, making it a
challenging task to compute the rates theoretically. The calculations of weak rates and
GT-strength distributions based on shell-model for pf- and sd-shell nuclei were performed by two
groups,~\citet{LangMar03} and~\citet{Oda94}, respectively. Their calculations took contributions
from ground as well as excited states.
To date, proton-neutron (pn-) quasi-particle random phase approximation (QRPA)~\citep{Nabi99a, Nabi04}
and the shell model~\citep{Lang00,LangMar03}, known as large-scale shell model (LSSM)
are considered as
most reliable models for the computations
 of weak rates on microscopic basis. However, the methodology of
shell model is based on Brink's hypothesis~\citep{BAH} for the inclusion of GT excited states
distributions. While, state-by-state calculations of GT-distributions from excited states are
performed in the pn-QRPA model. In addition, this model involves a luxuriant model space extending up to
7$\hbar\omega$ which can handle arbitrarily massive nuclei. Nabi and collaborators
successfully employed this model to study beta-decay properties of several Fe-peak nuclei (e.g., see~\citep{shehzadi20, Nabi19, Majid18, Nabi17, Muneeb13, Nabi08, Nabi07, Nabi05}).

GT-transitions of medium-mass nuclei (46 $<$ A $<$ 70) are of particular significance in latter
stages of nuclear matter evolution of stars. Amongst these nuclei, weak-rates due to Scandium isotopes have central
importance during pre-collapse stages of high-mass stars, where value of Y$_{e}$ changes from
0.40 to 0.50~\citep{Aufder94b, Heger01}. In this work, we will focus on the weak electron capture (EC) and $\beta^{-}$-decay reactions due to isotopes of scandium (Equation~\ref{Eq:weakdecays}) and will calculate their corresponding rates.
\begin{eqnarray}
\beta^{-}\text{decay}: &^{A}_{Z}\text{Sc} \rightarrow _{Z+1}^{A}\text{Ti} + e^{-} + \bar{\nu}&  \nonumber \\
\text{EC}: &^{A}_{Z}\text{Sc} + e^{-} \rightarrow _{Z-1}^{A}\text{Ca} + \nu&.
\label{Eq:weakdecays}
\end{eqnarray}
In the simulation studies of Aufderheide and collaborators,
the reported astrophysically important $\beta^{-}$-decay and EC isotopes of scandium involve $^{50-54}$Sc and
$^{48-51}$Sc, respectively. In a recent study, \citet{Nabi21} reported a detailed
investigation of most important presupernova nuclei including some Sc-isotopes ($^{48,49,50,51,52}$Sc)
based on the pn-QRPA model. They computed the mass fraction, nuclear partition functions, weak rates and
temporal rate of change of Y$_{e}$. In their study, the authors listed EC and $\beta^{-}$-decay Sc-nuclei
$^{48,49,50}$Sc which influence Y$_{e}$ greatly after silicon burning stage of core.
Several other studies on different Sc-isotopes also revealed their astrophysical implication. These include shell model study of $^{48}$Sc by~\citet{Caurier94}, $^{47,49}$Sc by~\citet{Mart97} and $^{50-52}$Sc by~\citet{Poves01}, with the consideration of full pf-shell space.
In their work, they computed energy spectra, static moments, electromagnetic transitions and
$\beta$-decay properties. In some recent studies, authors focused on many medium-mass Sc-isotopes.
The $\beta$-decay characteristics and ground-state properties of Sc-isotopes having N $>$ 28
have been studied by~\citet{Borzov18} using DF+CQRPA model. In~\citet{Possidonio18}, the authors
calculated the $\beta$-decay rates of Sc-isotopes (46 $\leq$ A $\leq$ 60) using Gross theory.
They studied the effect of different values of axial-vector coupling constant and different
energy distribution functions on their rates. Later, within the Gross theory,~\citet{Azevedo20a}
analysed the effect of anti-neutrino mass on $\beta^{-}$-decay rates and
calculated decay rates for several Sc-isotopes~\citep{Azevedo20b}.

In our present study, we have computed the terrestrial $\beta$-decay half-lives (T$_{1/2}$) of
$^{49-54}$Sc nuclei. We have also done the comparison of our pn-QRPA T$_{1/2}$
with experimental~\citep{Aud17} data and different theoretical models.
In addition, we have determined the B(GT) strengths and rates of $\beta^{-}$-decay and EC reactions. Our computed rates have also been compared with the calculations of earlier rates
based on LSSM and IPM models. The brief description of the formalism of pn-QRPA model which is adopted in current work is presented
in next Section. In Section~\ref{sec:results} we have reported our results and discussed them.
At the end, we have concluded our results in Section~\ref{sec:conclusions}.

\section{Formalism}
\label{sec:formalism}

The system of quasi-particles (qp) in the model of pn-QRPA is treated as single-particle
(sp) states. The interactions between these qp are included through correlated pairing forces between
nucleons and residual GT-forces between proton-neutron pairs.
The Hamiltonian of this system is written as
\begin{equation}
H_{qrpa} = h_{sp} + \nu_{pairing} + \nu_{pp(GT)} + \nu_{ph(GT)}~,
\label{Ham}
\end{equation}
where $h_{sp}$ corresponds to the $sp$-Hamiltonian. The Nilsson model~\citep{Nil55} is employed
to determine the energies and state functions of sp-system. The second term $\nu_{pairing}$
corresponds to the nucleon-nucleon pairing interaction based on the BCS approximation.
The inclusion of GT-forces with particle-particle ($pp$) and particle-hole ($ph$)
matrix elements were employed through last two terms, $\nu_{pp(GT)}$ and $\nu_{ph(GT)}$, respectively.
The constants $\kappa$ and $\chi$ were introduced for the pp and ph GT-interactions,
respectively. These constant values were fine tuned to produce the $\beta$-decay half-lives
consistent with experimentally observed data of~\citet{Aud17}. The $\kappa$ and $\chi$ were parameterized
in accordance with 1/A$^{0.7}$ dependence~\citep{Hom96} and given by
\begin{equation*}
\chi = 8.54/A^{0.7} \;(MeV);~~~~ \kappa = 1.525/A^{0.7} \;(MeV)~.
\end{equation*}
The explicit values of the optimized $\chi$ and $\kappa$ parameters
are also shown in Table~\ref{table1}. Other parameters integrated
into the pn-QRPA are the pairing gaps ($\Delta _{p(n)}$), the
nuclear deformation parameter ($\varepsilon_{2}$), Q-values, the
Nilsson potential (NP) parameters taken from~\citet{Rag84} and the
Nilsson oscillator constant (Equation~\ref{OC}). For the pairing
gaps, we used globally accepted expression,
\begin{equation}
\Delta _{p} = \Delta _{n} = 12A^{-1/2}\;(MeV)~.
\label{PG}
\end{equation}
\begin{equation}
\hbar\omega = 41A^{-1/3}\;(MeV)~.
\label{OC}
\end{equation}
The $\varepsilon_{2}$ were calculated by
\begin{eqnarray}
\varepsilon_{2} = \frac{RQ}{A^{2/3}Z};~~~~R=\frac{125}{1.44}~.
\label{Eq:Dpar}
\end{eqnarray}
The Equation~\ref{Eq:Dpar} depends on the electric quadruple moment ($Q$),
Z (atomic number) and A (mass number). In the present calculations, we used the
values of $Q$ as reported in~\cite{Mol81}.
The experimental values of mass excess
form~\cite{Aud17} were used to compute the Q-values for corresponding decay transitions.

The decay rates for EC and $\beta^{-}$-transitions, from parent $i^{th}$ state to the
daughter nucleus $j^{th}$ state under stellar conditions were computed by
\begin{eqnarray}
\lambda ^{\beta^{-}(EC)} _{ij} &=& \ln 2 \frac{f_{ij}^{\beta^{-}(EC)} (\rho, E_{f}, T)}{D/B_{ij}},
\label{Eq:rate}
\end{eqnarray}
where constant D = 6143 (adopted from~\cite{Har09}) and reduced probability of
$\beta^{-}$ (EC) transition is
\begin{eqnarray}
B_{ij} = \frac{B(GT)_{ij}}{\left(g_{A}/ g_{V}\right)^{-2}} + B(F)_{ij},
\label{Eq:Rprob}
\end{eqnarray}
which incorporates the sum of probabilities of Fermi (B(F)$_{ij}$) and Gamow-Teller (B(GT)$_{ij}$)
transitions. These transitions were computed by Equations~\ref{Eq:FTP} and~\ref{Eq:GTP}, respectively.
The value of $g_{A}/g_{V}$ = -1.2694 (reported in~\cite{Nak10}).

\begin{eqnarray}
B(F)_{ij} &=& \frac{|\langle j||\sum_{l}t^{l}_{\pm}||i \rangle|^{2}}{2J_{i}+1}~,
\label{Eq:FTP}
\end{eqnarray}
\begin{eqnarray}
B(GT)_{ij} &=& \frac{|\langle j||\sum_{l}t^{l}_{\pm}\vec{\sigma}^{l}||i \rangle|^{2}}{2J_{i}+1}~,
\label{Eq:GTP}
\end{eqnarray}
where $J_{i}$ is the total spin of the nucleus for $i^{th}$ state and
$t^{l}$ and $\vec{\sigma}^{l}$ are the  isospin (raising/lowering) and
spin operators, respectively.

In Equation \eqref{Eq:rate}, the phase space integrals (in natural units)
for $\beta^{-}$-decay ($f_{ij}^{\beta^{-}}$) and EC ($f_{ij}^{EC}$) are
\begin{eqnarray}
f_{ij}^{\beta^{-}} &=& \int_{1}^{\mathcal{w}_{\mathcal{m}}}\mathcal{w}
(\mathcal{w}_{\mathcal{m}}-\mathcal{w})^{2}(\mathcal{w}^{2} -1)^{1/2}\\ \nonumber
 &\times& F(+Z,\mathcal{w})(1-Z_{-})d\mathcal{w}~,
\label{Eq:FIBeta}
\end{eqnarray}
\begin{equation}
f_{ij}^{EC} = \int_{\mathcal{w}_{l} }^{\infty }\mathcal{w}
(\mathcal{w}_{m} +\mathcal{w})^{2}(\mathcal{w}^{2} -1)^{1/2} F(+
Z, \mathcal{w}) Z_{-}d\mathcal{w}~,
\label{Eq:FIEC}
\end{equation}
where $\mathcal{w}$ incorporates the rest mass energy and K.E. of electron,
$\mathcal{w}_{m}$ ($\mathcal{w}_{l}$) total energy for $\beta^{-}$-transition (total threshold
energy for EC decay) and Z$_{-}$ is the electron distribution function obeying
Fermi-Dirac statistics. The Fermi function  $F\left(+Z,\mathcal{w}\right)$ were
calculated by using~\cite{Gove71} method. The total $\beta^{-}$-decay (EC)
rates for a nucleus were calculated using
\begin{equation}
\lambda^{\beta^{-}(EC)} = \sum _{ij}P_{i} \lambda _{ij}^{\beta^{-}(EC)}~,
\label{Trate}
\end{equation}
where $P_{i}$ corresponds to occupation probability of excitation levels in parent
nucleus obeying the normal Boltzmann distribution.
Assuming thermal equilibrium, the probability of occupation of parent excited state $i$ was estimated using
\begin{equation}
 P_{i} = \frac{\exp({\frac{-E_{i}}{KT}})}{\sum_{i=1}\exp({\frac{-E_{i}}{KT}})}.
\label{prob}
\end{equation}
The summation in Equation
\eqref{Trate} was taken over all initial and final levels until reasonable convergence was obtained in the calculated rates.

\section{Results and discussions}
\label{sec:results}

In this study, we have evaluated the GT-strength distributions of some selected
unstable isotopes of Scandium, $^{49-54}$Sc, by using the pn-QRPA model. In addition,
the allowed weak $\beta^{-}$-decay \& EC rates and terrestrial half-lives have been estimated.
The calculations of weak rates have been performed at stellar temperatures and density covering the
ranges ($10^{7} - 3 \times 10^{10}$)\;K and ($10-10^{11}$)\;g/cm$^{3}$, respectively. We present
the comparison of presently calculated rates with the earlier rates of IPM and LSSM,
where available. Since, it has been observed that, the experimentally measured
GT-strengths are generally smaller in magnitude than the calculations of nuclear models. So, the calculated GT strengths are usually renormalized by using some fixed quenching factor by different models. For RPA calculations, a standard quenching factor of 0.6 is mostly used, e.g. in~\citep{Vetterli89, Ron93, Gaarde83} and hence is also employed in the current calculations.

\begin{figure}[h]
\begin{center}
\includegraphics[width=0.7\textwidth]{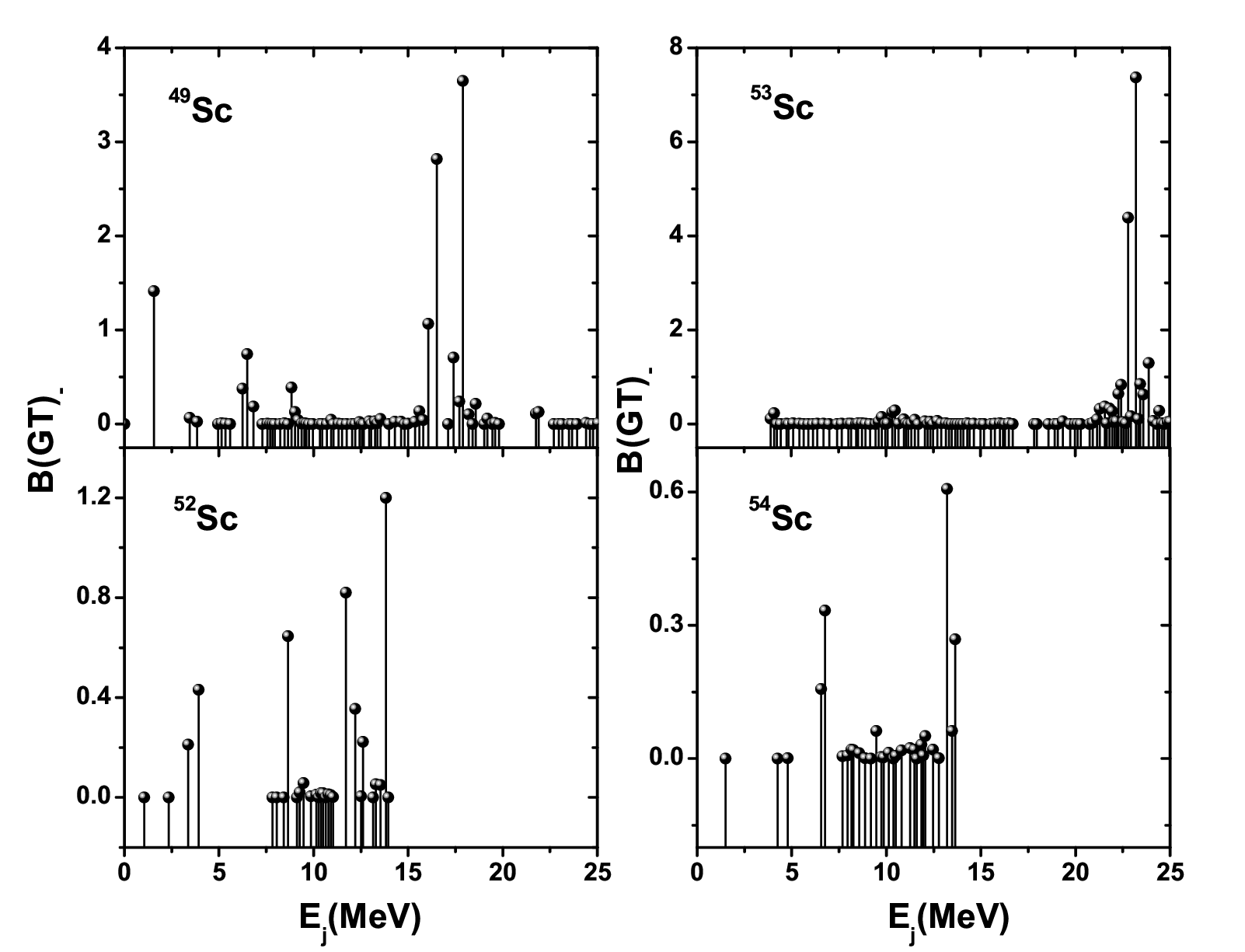}
\vspace{-0.1cm}\caption{The GT-strength distributions B(GT)$_{-}$
for $^{49, 52, 53, 54}$Sc as a function of E$_{j}$ (daughter
excitation energies in MeV units) using the pn-QRPA model.}
\label{figure1}
\end{center}
\end{figure}

\begin{figure}[h]
\begin{center}
\includegraphics[width=0.7\textwidth]{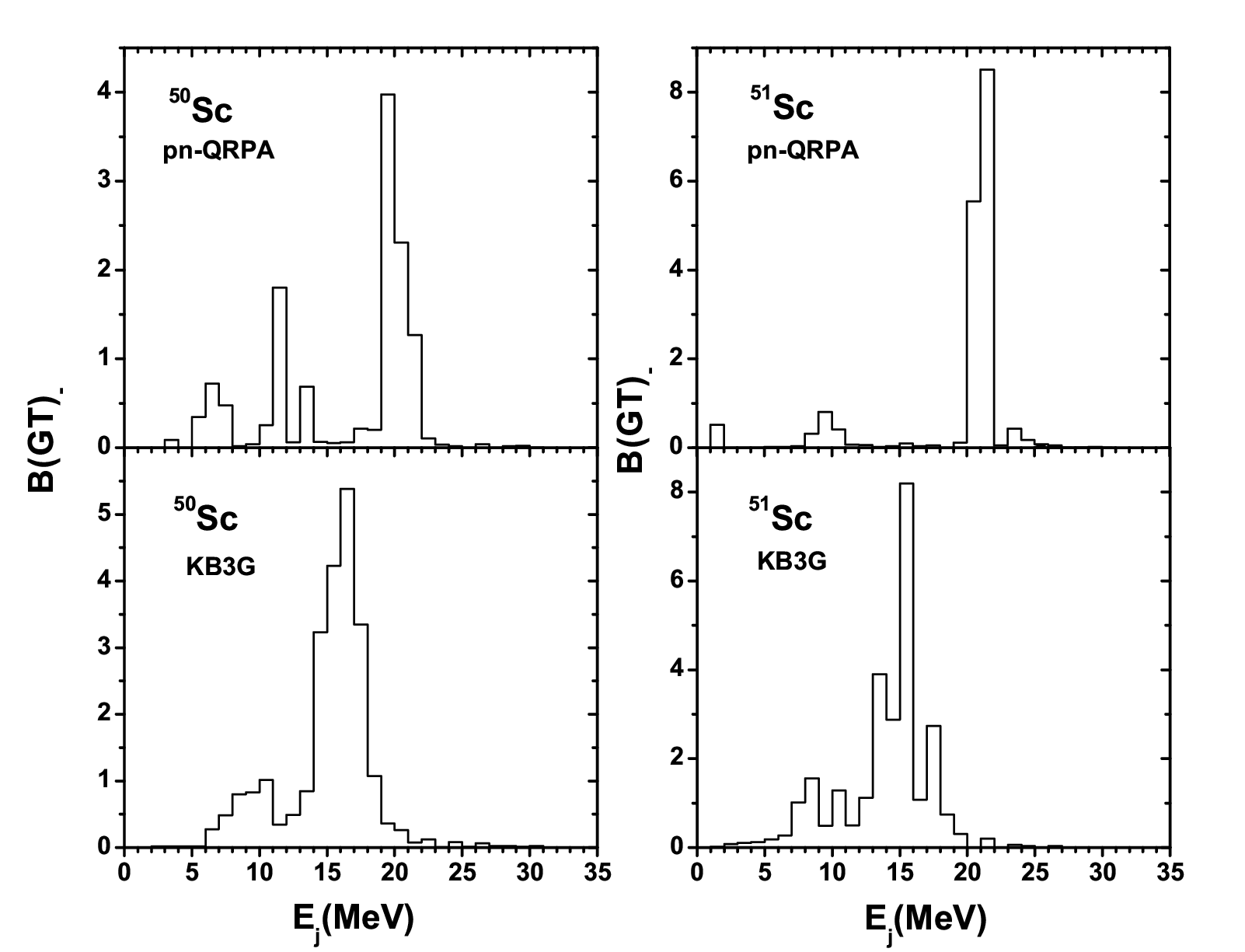}
\vspace{-0.1cm}\caption{The comparison of calculated B(GT)$_{-}$
strength distributions in $^{50, 51}$Sc with shell-model
calculation~\citep{Poves01} based on KB3G effective interaction. The
B(GT)$_{-}$ values are summed up in MeV bins.} \label{figure2}
\end{center}
\end{figure}

The comparison of our calculated half-lives under terrestrial
conditions with other theoretical model calculations and
experimental data~\citep{Aud17}, has been shown in
Table~\ref{table2}. A good agreement of our pn-QRPA calculated
T$_{1/2}$ values with the corresponding experimental data and the
shell-model calculations done using KB3G mentioned
in~\citep{Poves01} and using KB3 in~\citep{Mart97} can be seen from
Table~\ref{table2}. However, other theoretical calculations of
T$_{1/2}$ of Sc isotopes done by~\cite{Borzov18} using DF+CQRPA,
\cite{Possidonio18} using Gross Theory and ~\cite{Mol2019} show
differences to the experimental data. The calculations of
GT-strengths in $\beta^{-}$ direction (B(GT)$_{-}$) are done for
ground state as well as for excited states of $^{49-54}$Sc isotopes.
However, due to space consideration, only ground state B(GT)$_{-}$
have been shown here. The electronic files of these strengths may be
requested from corresponding author. The GT-strength distributions
with respect to ground states of corresponding parent $^{49, 52, 53,
54}$Sc nuclei along $\beta^{-}$-decay direction are shown in
Figure~\ref{figure1}, where B(GT)$_{-}$ are taken along ordinate as
a function of excitation energies (E$_{j}$) of titanium daughter
isotopes. This figure depicts that the B(GT)$_{-}$ are highly
fragmented over daughter nuclei states. Figure~\ref{figure2} shows
the comparison of our calculated B(GT)$_{-}$ strengths for $^{50,
51}$Sc isotopes (upper panels) with those from
shell-model~\citep{Poves01} calculations (lower panels) where the
authors used KB3G effective interaction for the calculations. For
the sake of comparison with~\citet{Poves01} data, for these two
isotopes, BGT strengths are summed up in MeV bins. This comparison
shows that in our calculations, for both of these nuclei, the peak
of GT-strength is obtained at higher excitation energies as compared
to the shell-model calculations.

\begin{figure}[h]
\begin{center}
\includegraphics[width=0.7\textwidth]{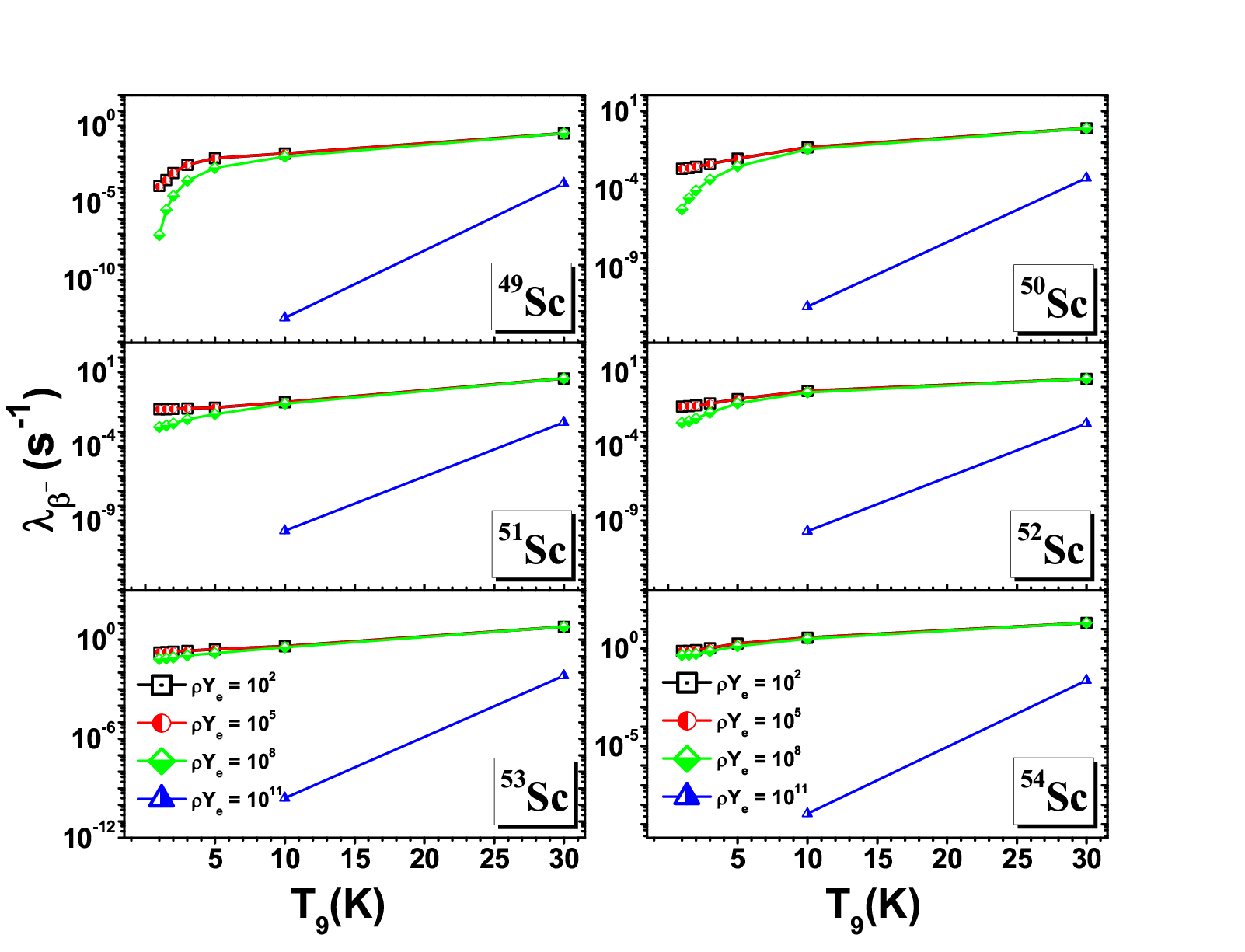}
\vspace{-0.1cm}\caption{The pn-QRPA calculated $\beta^{-}$-decay
rates for scandium isotopes at different stellar densities as a
function of temperature. Temperatures (T$_{9}$) are given in units
of $10^{9}\;$K. $\rho$Y$_{e}$ has units of g$\;$cm$^{-3}$, where
$\rho$ is the baryon density and Y$_{e}$ is lepton-to-baryon
fraction.} \label{figure3}
\end{center}
\end{figure}

\begin{figure}
\begin{center}
\includegraphics[width=0.8\textwidth]{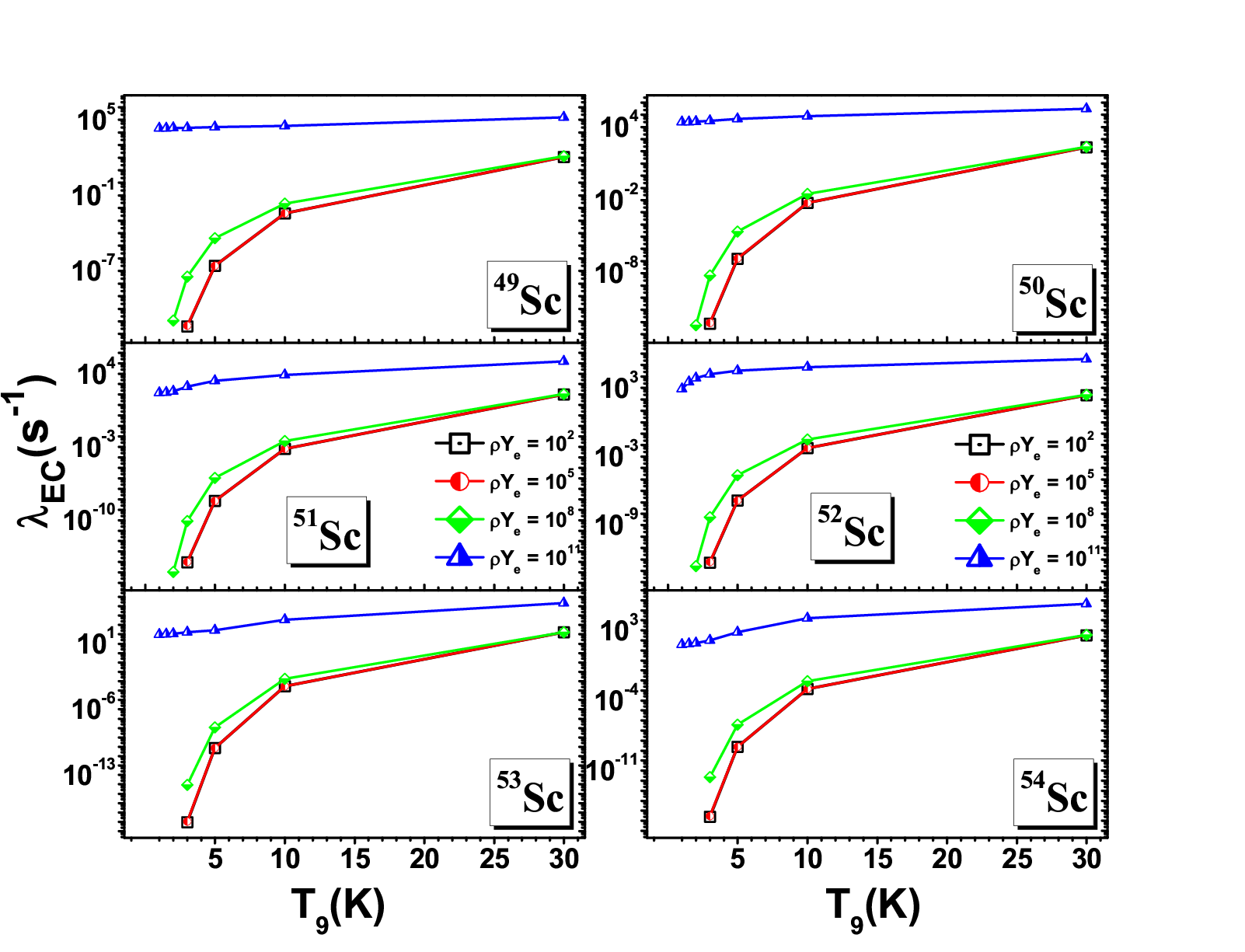}
\vspace{-0.1cm}\caption{The pn-QRPA calculated electron capture (EC)
rates for scandium isotopes at different stellar densities as a
function of temperature. Other details same as in
Figure~\ref{figure3}. } \label{figure4}
\end{center}
\end{figure}

Next, we analyse the results of our computed rates of
$\beta^{-}$-decay and EC reactions over stellar domain.
Figure~\ref{figure3} depicts the $\beta^{-}$-decay rates on
$^{49-54}$Sc. Similarly, EC rates for these isotopes are shown in
Figure~\ref{figure4}. We present the weak rates over a wide
temperature (T$_{9}$) domain in the units of 10$^{9}$ K. The rates
are presented in logarithmic scale with base 10 (log$\lambda
_{EC(\beta^{-})}$) in units of s$^{-1}$ at low
($\rho$Y$_{e}$=10$^{2}$\;g/cm$^{3}$), medium
($\rho$Y$_{e}$=10$^{5}$, 10$^{8}$\;g/cm$^{3}$) and high
($\rho$Y$_{e}$=10$^{11}$\;g/cm$^{3}$) densities.
Figure~\ref{figure3} shows that for each isotope under study,
$\beta^{-}$-decay rates enhance with increment in temperature of
core material. The rate of increment is higher and prominent at high
density. With rise in temperature the increase in $\beta^{-}$-decay
rates happens because available phase space expand largely with
temperature. In addition, the contributions of partial rates to the
total rates increase due to the increase in occupation probabilities
of parent excited states with temperature rise. With the increase in
density to around 10$^{5}$\;g/cm$^{3}$, the $\beta^{-}$-decay rates
remain nearly constant. However, as the core material gets further
dense, the rates decrease by several orders of magnitude, especially
in low and medium T$_{9}$ regions. This happens due to the reduction
in available phase space when core material stiffens.
Figure~\ref{figure4} shows that as both the stellar density and
temperature increase, the EC rates get enhanced. The rate of
increment in EC rates in low and medium density (temperature)
regions is several orders of magnitude larger in comparison to other
physical conditions. The reason for this increase in rates is the
electron Fermi energy which goes to higher level and more parent
excited states are occupied as the core gets stiffens and its
temperature rises, respectively. However, around high density
10$^{11}$\;g/cm$^{3}$ EC rates are almost constant.

The comparison of our calculated rates of $\beta^{-}$-decay (EC)
with the previously computed rates based on IPM and LSSM is
presented in Table~\ref{table3} (Table~\ref{table4}). Due to space
consideration, comparison has been shown on selected values of
stellar temperature (T$_{9}$) and density ($\rho$Y$_{e}$). First and
second columns of both tables show these values of $\rho$Y$_{e}$ and
T$_{9}$, respectively. Remaining columns of Table~\ref{table3}
(Table~\ref{table4}) depict the comparison ratios of our rates for
$\beta^{-}$-decay (EC) to that of IPM and LSSM, separately. In case
of $^{49,50,51}$Sc, the IPM rates are in general $\sim$1-3 orders of
magnitude greater than our $\beta^{-}$-decay rates, especially at
high temperature and density. The IPM results were obtained without
considering quenched GT-strengths. Further, in IPM, the process of
particle emission from excited states was not taken into effect and
their parent excitation energies extended well beyond the particle
decay channel. At high temperatures and densities, the probability
for the occupation of high-lying excited states become finite and
their contributions begin to show their cumulative effect. Whereas,
in our pn-QRPA based formalism, the GT-strengths of all excited
states in parent nuclei were computed in microscopical
state-by-state way. These facts could be the cause of enhancement in
IPM rates. In low temperature (T$_{9} = 1, 1.5, 2\;$K) and density
($\rho$Y$_{e}$=10$^{2}$, 10$^{5}$\;g/cm$^{3}$) regions, our
$\beta^{-}$-decay rates are roughly equal to IPM rates. At medium
temperatures (T$_{9} = 3, 5\;$K), for $\rho$Y$_{e}\leq$
10$^{8}$\;g/cm$^{3}$ our rates on $^{49}$Sc are enhanced by $\sim$ 1
order of magnitude. This unusual enhancement in our rates may occurs
due to the unmeasured matrix elements in IPM calculations which were
given too small approximated values. Like IPM, LSSM rates are in
general larger as compared to our $\beta^{-}$-decay rates for
$^{49-52}$Sc. A rough agreement between both model rates can be seen
in low and medium density regions ($\rho$Y$_{e}$ = 10$^{2}$,
10$^{5}$, 10$^{8}$\;g/cm$^{3}$). In addition, our present rates on
$^{51}$Sc are roughly equal to LSSM rates at $\rho$Y$_{e}$ =
10$^{11}$\;g/cm$^{3}$. Only in case of $^{49}$Sc, our rates at
T$_{9} = 3, 5\;$K are enhanced by about one order of magnitude.
Apparently, the same formalism is adopted for the calculations of
phase-space and Q-values in both, the pn-QRPA model and LSSM.
However, in LSSM theory the consideration of back-resonance and
Brink's hypothesis may cause the differences between the results of
both models.

Lastly, we move to the comparison of our calculated rates of EC
reactions with the corresponding LSSM and IPM rates.
Table~\ref{table4} shows that at most of the stellar density and
temperature conditions, our rates are larger than IPM and LSSM EC
rates. In case of $^{50}$Sc, at T$_{9}\geq 5\;$K, our rates for EC
are 1-2 orders of magnitude greater than both IPM and LSSM rates .
For temperature range 2 $\leq$ T$_{9}\leq 3\;$K, our rates are
roughly equal to the EC rates of IPM. Our rates on $^{49}$Sc are
enhanced than IPM rates by up to 2 orders of magnitude. Similarly,
in $^{49,51}$Sc, LSSM rates are smaller than our calculated EC
rates. Except, in $^{51}$Sc, at $\rho$Y$_{e}$ =
10$^{11}$\;g/cm$^{3}$, our rates are roughly equal to LSSM rates.
The reasons for the differences in our calculated results and the
LSSM/IPM results were discussed earlier.
\section{Conclusions}
\label{sec:conclusions}

A better understanding of mechanism involved in pre- and during supernova
explosions may rely on the reliable estimations of rates for weak EC and $\beta^{-}$-decay processes.
In addition, the contest between these rates with the change in density and temperature of the
stellar core may provide a deeper understanding of the processes associated with
the nucleosynthsis yield.

In general, when core gets hotter, both stellar $\beta^{-}$-decay and EC rates increase.
However, at low density, $\beta^{-}$-decay compete well with EC rates. As the core
gets denser, $\beta^{-}$-decay rates become insignificant as compared to
EC rates due to Pauli blockage of phase-space~\citep{lan01}. Thus, EC rates dominate
the overall scenario of late stages of core-collapse. In case of our present study
the EC and $\beta^{-}$-decay rates on $^{49-54}$Sc nuclei exhibits the same trend
with variation in stellar core temperature and density. In addition, our
calculated terrestrial half-lives of these nuclei show a good agreement
with measured experimental data~\citep{Aud17}.

We displayed the comparison of our stellar EC/$\beta^{-}$-decay rates
with that of LSSM and IPM. It is noted that our rates of $\beta^{-}$-decay reactions
are smaller than corresponding rates of both IPM and LSSM in the regions
of high stellar density and temperature. In contrast, enhancement in our EC rates
were noted as compared to IPM and LSSM rates in high temperature regions. However, in low
temperature (T$_{9}<$ 10 K) and high density ($\rho$Y$_{e}$ = 10$^{11}$\;g/cm$^{3}$)
regions, our EC rates have a rough agreement with IPM rates. The reason for
differences between our and their rates may be attributed to the application of
the Brink's hypothesis in their calculations. In addition, unquenched GT-strength, approximate
values of unmeasured matrix elements and computations of GT centroids by using
0$\hbar \omega$ shell model in IPM and employment of back-resonances in LSSM may contribute
to these differences. The present study of $^{49-54}$Sc isotopes may assist
in modeling the late stellar-evolutionary stages before going supernova in a more reliable fashion.

\clearpage \onecolumn

\begin{table}[pt]
\caption{\small Optimized values of $\chi$ and $\kappa$ for
$^{49-54}$Sc isotopes calculated by the pn-QRPA
model.}\label{table1} \centering { \scalebox{1.2}{
\begin{tabular}{ccc}
\toprule
Isotopes Sc & $\chi$ & $\kappa$ \\
(A) & (MeV) & (MeV)\\
\hline
\hline

49 & 0.560  & 0.100 \tabularnewline
50 & 0.552 & 0.099  \tabularnewline
51 & 0.545 & 0.097  \tabularnewline
52 & 0.537 & 0.096  \tabularnewline
53 & 0.530 & 0.095  \tabularnewline
54 & 0.523 & 0.093  \tabularnewline

 \bottomrule
\end{tabular}}}
\end{table}
\begin{table}[pt]
\caption{\small The comparison of pn-QRPA calculated terrestrial
half-lives of Sc isotopes with experimental data~\citep{Aud17} and
theoretical values taken from~\citep{Mol2019}, shell model
calculations done using KB3 mentioned in~\citep{Mart97}, using KB3G
in~\citep{Poves01}, DF+CQRPA~\citep{Borzov18} and Gross
Theory~\citep{Possidonio18}.}\label{table2} \centering {
\hspace{-1cm} {\small
\begin{tabular}{cccccccc}
& & & & & & & \\
\toprule
\multirow{2}{*}{Nuclei} & \multicolumn{7}{c} {Half-life (seconds)} \\
\cmidrule{2-3} \cmidrule{3-4}  \cmidrule{5-6}  \cmidrule{7-8} &
\multicolumn{1}{c}{pn-QRPA} & \multicolumn{1}{c}{Exp.} &
\multicolumn{1}{c}{M\"{o}ller} & \multicolumn{1}{c}{KB3G} &
\multicolumn{1}{c}{KB3} &
\multicolumn{1}{c}{DF+CQRPA}  & \multicolumn{1}{c}{Gross Theory}\\
\midrule
 $^{49}$Sc  &   3522.29  &  3430.80$\pm$7.8  &  102  &  --  &  2484  &  --  &  --  \\
$^{50}$Sc  &   102.52  &  102.50 $\pm$0.5  &  44.70  &  120$^{+2}_{-1}$  &  --  &  --  &  4.35  \\
$^{51}$Sc  &   12.41  &  12.40$\pm$0.1  &  2.95  &  12.30$\pm$0.3  &  --  &  --  &  11.06  \\
$^{52}$Sc  &   8.53  &  8.20$\pm$0.2  &  3.23  &  6.20$^{+1}_{-0.8}$ &  --  &  --  &  1.75  \\
$^{53}$Sc  &   2.44  &  2.40$\pm$0.6  &  0.13  &  --  &  --  &  5.53  &  2.36  \\
$^{54}$Sc  &   0.54  &  0.53$\pm$0.015  &  0.04  &  --  &  --  &  0.316  &  1.36  \\
 \bottomrule
\end{tabular}}}
\end{table}




\begin{table}[pt]
\caption{\small Ratios of the pn-QRPA calculated $\beta^{-}$-decay
rates due to scandium isotopes to those of calculated by large scale
shell model~\citep{Lang00} (R$_{\beta^{-}}$(LSSM)) and those
calculated by Independent particle model~\citep{Fuller}
(R$_{\beta^{-}}$(IPM)). The comparison has been shown at some
selected values of stellar densities and temperatures. Temperatures
(T$_{9}$) are given in units of $10^{9}\;$K. $\rho$Y$_{e}$ has units
of g$\;$cm$^{-3}$, where $\rho$ is the baryon density and Y$_{e}$ is
lepton-to-baryon fraction.}\label{table3} \centering {
\begin{tabular}{ccccccccc}
 & & & & & & & &\\
\toprule \multirow{2}{*}{$\rho$Y$_{e}$} & \multirow{2}{*}{T$_{9}$} &
\multicolumn{2}{c}{$^{49}$Sc}&
\multicolumn{2}{c}{$^{50}$Sc} & \multicolumn{2}{c}{$^{51}$Sc} & $^{52}$Sc \\
\cmidrule{3-4}  \cmidrule{5-6}  \cmidrule{7-8} \cmidrule{9-9} & &
\multicolumn{1}{c}{R$_{\beta^{-}}$(LSSM)} &
\multicolumn{1}{c}{R$_{\beta^{-}}$(IPM)} &
\multicolumn{1}{c}{R$_{\beta^{-}}$(LSSM)} &
\multicolumn{1}{c}{R$_{\beta^{-}}$(IPM)} &
\multicolumn{1}{c}{R$_{\beta^{-}}$(LSSM)} &
\multicolumn{1}{c}{R$_{\beta^{-}}$(IPM)} &
\multicolumn{1}{c}{R$_{\beta^{-}}$(LSSM)} \\
\midrule
$10^{2}$ & 1 & 6.38E-01 & 5.62E-01 & 2.68E-01 & 1.50E-01 & 5.65E-01 & 6.11E-01 & 5.48E-01\tabularnewline
$10^{2}$ & 1.5 & 1.59E+00 & 1.40E+00 & 2.54E-01 & 1.06E-01 & 5.68E-01 & 6.10E-01 & 5.90E-01\tabularnewline
$10^{2}$ & 2 & 4.36E+00 & 3.84E+00 & 2.65E-01 & 9.20E-02 & 5.77E-01 & 6.00E-01 & 6.62E-01\tabularnewline
$10^{2}$ & 3 & 1.49E+01 & 1.30E+01 & 3.28E-01 & 9.93E-02 & 6.17E-01 & 5.42E-01 & 8.57E-01\tabularnewline
$10^{2}$ & 5 & 3.59E+01 & 2.41E+01 & 3.49E-01 & 1.58E-01 & 6.73E-01 & 4.06E-01 & 9.95E-01\tabularnewline
$10^{2}$ & 10 & 1.57E+00 & 9.86E-01 & 2.85E-01 & 1.49E-01 & 4.88E-01 & 1.71E-01 & 8.09E-01\tabularnewline
$10^{2}$ & 30 & 2.03E+00 & 3.07E-02 & 1.19E+00 & 4.56E-02 & 5.36E+00 & 7.96E-02 & 2.59E+00\tabularnewline
 &  &  &  &  &  &  &  & \tabularnewline
$10^{5}$ & 1 & 5.96E-01 & 5.25E-01 & 2.65E-01 & 1.48E-01 & 5.64E-01 & 6.10E-01 & 5.45E-01\tabularnewline
$10^{5}$ & 1.5 & 1.57E+00 & 1.38E+00 & 2.52E-01 & 1.05E-01 & 5.65E-01 & 6.08E-01 & 5.89E-01\tabularnewline
$10^{5}$ & 2 & 4.35E+00 & 3.83E+00 & 2.64E-01 & 9.14E-02 & 5.74E-01 & 5.97E-01 & 6.61E-01\tabularnewline
$10^{5}$ & 3 & 1.49E+01 & 1.30E+01 & 3.27E-01 & 9.91E-02 & 6.17E-01 & 5.42E-01 & 8.55E-01\tabularnewline
$10^{5}$ & 5 & 3.58E+01 & 2.41E+01 & 3.49E-01 & 1.58E-01 & 6.73E-01 & 4.06E-01 & 9.95E-01\tabularnewline
$10^{5}$ & 10 & 1.57E+00 & 9.86E-01 & 2.85E-01 & 1.49E-01 & 4.88E-01 & 1.71E-01 & 8.09E-01\tabularnewline
$10^{5}$ & 30 & 2.03E+00 & 3.07E-02 & 1.19E+00 & 4.56E-02 & 5.36E+00 & 7.96E-02 & 2.59E+00\tabularnewline
 &  &  &  &  &  &  &  & \tabularnewline
$10^{8}$ & 1 & 9.42E-02 & 9.27E-02 & 1.81E-03 & 9.93E-04 & 7.64E-02 & 9.86E-02 & 7.52E-02\tabularnewline
$10^{8}$ & 1.5 & 1.68E+00 & 1.41E+00 & 7.14E-03 & 2.74E-03 & 9.53E-02 & 1.21E-01 & 9.98E-02\tabularnewline
$10^{8}$ & 2 & 7.01E+00 & 6.21E+00 & 1.84E-02 & 5.86E-03 & 1.29E-01 & 1.52E-01 & 1.48E-01\tabularnewline
$10^{8}$ & 3 & 2.57E+01 & 2.36E+01 & 6.50E-02 & 1.92E-02 & 2.33E-01 & 2.09E-01 & 3.31E-01\tabularnewline
$10^{8}$ & 5 & 3.34E+01 & 1.80E+01 & 1.53E-01 & 8.07E-02 & 4.01E-01 & 2.24E-01 & 6.49E-01\tabularnewline
$10^{8}$ & 10 & 1.12E+00 & 7.59E-01 & 2.38E-01 & 1.24E-01 & 4.34E-01 & 1.49E-01 & 7.18E-01\tabularnewline
$10^{8}$ & 30 & 2.01E+00 & 3.02E-02 & 1.19E+00 & 4.50E-02 & 5.35E+00 & 7.93E-02 & 2.57E+00\tabularnewline
 &  &  &  &  &  &  &  & \tabularnewline
$10^{11}$ & 10 & 6.17E-02 & 2.77E-03 & 1.95E-02 & 2.25E-03 & 4.10E-01 & 2.25E-02 & 6.04E-02\tabularnewline
$10^{11}$ & 30 & 9.68E-01 & 6.30E-03 & 5.93E-01 & 1.73E-02 & 3.48E+00 & 3.44E-02 & 1.36E+00\tabularnewline
 \bottomrule
\end{tabular}}
\end{table}

\begin{table}[pt]
\caption{\small Ratios of the pn-QRPA calculated electron capture
rates due to scandium isotopes to those of calculated by large scale
shell model~\citep{Lang00} (R$_{\text{EC}}$(LSSM)) and those
calculated by Independent particle model~\citep{Fuller}
(R$_{\text{EC}}$(IPM)). Other details same as in
Table~\ref{table3}.}\label{table4} \centering {
\begin{tabular}{ccccccc}
 & & & & & & \\
\toprule \multirow{2}{*}{$\rho$Y$_{e}$} & \multirow{2}{*}{T$_{9}$} &
\multicolumn{2}{c}{$^{49}$Sc}&
\multicolumn{2}{c}{$^{50}$Sc} & $^{51}$Sc \\
\cmidrule{3-4}  \cmidrule{5-6}  \cmidrule{7-7} & &
\multicolumn{1}{c}{R$_{\text{EC}}$(LSSM)} &
\multicolumn{1}{c}{R$_{\text{EC}}$(IPM)} &
\multicolumn{1}{c}{R$_{\text{EC}}$(LSSM)} &
\multicolumn{1}{c}{R$_{\text{EC}}$(IPM)} &
\multicolumn{1}{c}{R$_{\text{EC}}$(LSSM)} \\
\midrule
$10^{2}$ & 3 & 1.05E+02 & 1.52E+02 & 6.95E+00 & 8.81E+00 & 7.19E+01\tabularnewline
$10^{2}$ & 5 & 7.66E+01 & 1.54E+02 & 4.15E+01 & 6.00E+01 & 4.92E+01\tabularnewline
$10^{2}$ & 10 & 3.55E+01 & 4.67E+01 & 9.27E+01 & 1.34E+02 & 3.21E+01\tabularnewline
$10^{2}$ & 30 & 2.75E+01 & 1.34E+01 & 7.66E+01 & 4.90E+01 & 4.67E+01\tabularnewline
 &  &  &  &  &  & \tabularnewline
$10^{5}$ & 3 & 1.05E+02 & 1.52E+02 & 6.93E+00 & 8.79E+00 & 7.19E+01\tabularnewline
$10^{5}$ & 5 & 7.66E+01 & 1.54E+02 & 4.15E+01 & 6.00E+01 & 4.92E+01\tabularnewline
$10^{5}$ & 10 & 3.55E+01 & 4.67E+01 & 9.27E+01 & 1.34E+02 & 3.21E+01\tabularnewline
$10^{5}$ & 30 & 2.75E+01 & 1.34E+01 & 7.66E+01 & 4.90E+01 & 4.67E+01\tabularnewline
 &  &  &  &  &  & \tabularnewline
$10^{8}$ & 2 & 1.95E+02 & 2.16E+02 & 8.69E-01 & 7.96E-01 & 1.66E+02\tabularnewline
$10^{8}$ & 3 & 2.04E+02 & 2.28E+02 & 9.38E+00 & 8.77E+00 & 9.98E+01\tabularnewline
$10^{8}$ & 5 & 1.22E+02 & 1.87E+02 & 5.21E+01 & 6.00E+01 & 5.83E+01\tabularnewline
$10^{8}$ & 10 & 3.86E+01 & 4.73E+01 & 9.84E+01 & 1.35E+02 & 3.36E+01\tabularnewline
$10^{8}$ & 30 & 2.76E+01 & 1.34E+01 & 7.67E+01 & 4.90E+01 & 4.68E+01\tabularnewline
 &  &  &  &  &  & \tabularnewline
$10^{11}$ & 1 & 1.45E+01 & 3.59E+00 & 1.85E+01 & 4.95E+00 & 1.61E-01\tabularnewline
$10^{11}$ & 1.5 & 1.45E+01 & 3.61E+00 & 1.93E+01 & 5.12E+00 & 1.73E-01\tabularnewline
$10^{11}$ & 2 & 1.47E+01 & 3.64E+00 & 2.07E+01 & 5.42E+00 & 2.33E-01\tabularnewline
$10^{11}$ & 3 & 1.53E+01 & 3.81E+00 & 2.49E+01 & 6.41E+00 & 6.25E-01\tabularnewline
$10^{11}$ & 5 & 1.71E+01 & 4.26E+00 & 3.52E+01 & 8.95E+00 & 2.22E+00\tabularnewline
$10^{11}$ & 10 & 2.04E+01 & 5.08E+00 & 5.38E+01 & 1.39E+01 & 7.66E+00\tabularnewline
$10^{11}$ & 30 & 4.03E+01 & 1.40E+01 & 9.86E+01 & 3.70E+01 & 5.92E+01\tabularnewline
 \bottomrule
\end{tabular}}
\end{table}

\end{document}